\documentclass[lettersize,journal]{IEEEtran}
\usepackage[printonlyused]{acronym}
\usepackage[style=ieee,backend=bibtex]{biblatex}
\bibliography{biblio}
\usepackage{fancyhdr}
\usepackage{amsmath,amsfonts}
\usepackage{algorithmic}
\usepackage{algorithm}
\usepackage{array}
\usepackage{subfigure}
\usepackage{textcomp}
\usepackage{stfloats}
\usepackage{url}
\usepackage{verbatim}
\usepackage{graphicx}
\usepackage{xcolor}
\colorlet{red}{black}

\begin{document}

\title{Predictive Context-Awareness for Full-Immersive Multiuser Virtual Reality with Redirected Walking}

\author{Filip Lemic, Jakob Struye, Thomas Van Onsem, Jeroen Famaey, Xavier Costa-Pérez\vspace{-3mm}
        % <-this % stops a space
\thanks{F. Lemic and X. Costa-Pérez are affiliated with the i2Cat Foundation, Spain, email: \{name.surname@i2cat.net\}. X. Costa-Pérez is also affiliated with NEC Laboratories Europe and ICREA, Spain.}% <-this % stops a space
\thanks{J. Struye, T. Van Onsem, and J. Famaey are affiliated with the University of Antwerp, Belgium email; \{name.surname\}@uantwerpen.be. J. Struye and J. Famaey are also affiliated with IMEC, Belgium.}% <-this % stops a space
\thanks{Manuscript received October 31, 2022; revised XXX.}}

% \markboth{IEEE Communications Magazine,~Vol.~XX, No.~XX, November~2022}%
\markboth{IEEE DRAFT}%
{Shell \MakeLowercase{\textit{et al.}}: A Sample Article Using IEEEtran.cls for IEEE Journals}

\fancyhf{}
\renewcommand{\headrulewidth}{0pt}
\fancyfoot[c]{}
\fancypagestyle{FirstPage}{
\vspace{-3mm}
\lfoot{\footnotesize \copyright2023 IEEE. Personal use of this material is permitted. Permission from IEEE must be obtained for all other uses, in any current or future media, including reprinting/republishing this material for advertising or promotional purposes, creating new collective works, for resale or redistribution to servers or lists, or reuse of any copyrighted component of this work in other works.\hfill} 
}

\maketitle

\begin{abstract}
The advancement of \ac{VR} technology is focused on improving its immersiveness, supporting multiuser \acp{VE}, and enabling users to move freely within their \acp{VE} while remaining confined to specialized VR setups through Redirected Walking (RDW). 
To meet their extreme data-rate and latency requirements, future \ac{VR} systems will require supporting wireless networking infrastructures operating in \ac{mmWave} frequencies that leverage highly directional communication in both transmission and reception through beamforming and beamsteering. 
We propose the use of predictive context-awareness to optimize transmitter and receiver-side beamforming and beamsteering. 
By predicting users' short-term lateral movements in multiuser \ac{VR} setups with \ac{RDW}, transmitter-side beamforming and beamsteering can be optimized through \ac{LoS} ``tracking'' in the users' directions.
At the same time, predictions of short-term orientational movements can be utilized for receiver-side beamforming for coverage flexibility enhancements. 
We target two open problems in predicting these two context information instances: i) predicting lateral movements in multiuser \ac{VR} settings with \ac{RDW}, and ii) generating synthetic head rotation datasets for training orientational movements predictors. 
Our experimental results demonstrate that \acf{LSTM} networks feature promising accuracy in predicting lateral movements, and context-awareness stemming from \acp{VE} further enhances this accuracy. 
Additionally, we show that a TimeGAN-based approach for orientational data generation can create synthetic samples that closely match experimentally obtained ones.
\end{abstract}


\thispagestyle{FirstPage}
%!TEX root = ieee_commag_main.tex

\acrodef{IoT}{Internet of Things}
\acrodef{ISM}{Industrial, Scientific and Medical}
\acrodef{LPWAN}{Low-Power Wide Area Network}
\acrodef{REM}{Radio Environmental Map}
\acrodef{SNR}{Signal-to-Noise Ratio}
\acrodef{MT}{Mobile Terminal}
\acrodef{LoS}{Line-of-Sight}
\acrodef{NLoS}{Non-Line-of-Sight}
\acrodef{MCS}{Modulation and Coding Scheme}
\acrodef{GPS}{Global Positioning System}
\acrodef{RSSI}{Received Signal Strength Indication}
\acrodef{MAE}{Mean Absolute Error}
\acrodef{SE}{Squared Error}
\acrodef{SDR}{Short-Range Device}
\acrodef{QoE}{Quality of Experience}
\acrodef{VR}{Virtual Reality}
\acrodef{AP}{Access Point}
\acrodef{HMD}{Head-Mounted Device}
\acrodef{IRS}{Intelligent Reflective Surface}
\acrodef{AoA}{Angle of Arrival}
\acrodef{RNN}{Recurrent Neural Network}
\acrodef{LSTM}{Long Short-Term Memory}
\acrodef{GRU}{Gated Recurrent Unit}
\acrodef{MSE}{Mean-Square Error}
\acrodef{APF}{Artificial Potential Field}
\acrodef{APF-R}{Artificial Potential Field Resetting}
\acrodef{HD}{High Definition}
\acrodef{RNN}{Recurrent Neural Network}
\acrodef{VE}{Virtual Experience}
\acrodef{RDW}{Redirected Walking}
\acrodef{FFT}{Fast Fourier Transform}
\acrodef{DL}{Deep Learning}
\acrodef{PDF}{Probability Density Function}
\acrodef{GAN}{Generative Adversarial Network}
\acrodef{mmWave}{millimeter Wave}
\acrodef{3D}{3-dimensional}
%!TEX root = ieee_commag_main.tex
\section{Introduction}

% \acf{VR} is benefiting different communities by revolutionizing their digital perceptions and interactions~\cite{munoz2020augmented}. 
% VR setups and contents are being continuously improved, with the main research efforts targeting enhancements in the immersiveness of \acfp{VE}. 
% More precisely, the research is focusing on enhancing the quality of \acp{VE} that are being presented to the users~\cite{zhang2019wireless}, as well as  on ``cutting the wire'' and supporting their wireless delivery without users' mobility constraints~\cite{chen2018virtual}. 
% An additional objective is to enable multiuser experiences, in which the users are able to collaborate in a way that the action of one user in a \ac{VE} is delivered to the others~\cite{bachmann2019multi} and consequently affects their \acp{VE}. 

The utilization of \acf{VR} technology is transforming digital experiences and interactions of various communities~\cite{munoz2020augmented}. 
To improve the immersiveness of \acfp{VE}, \ac{VR} setups and content are continually being upgraded. 
Research efforts are primarily focused on enhancing the quality of \acp{VE} provided to the users~\cite{zhang2019wireless}, and facilitating its wireless delivery without mobility constraints, also known as ``cutting the wire''~\cite{chen2018virtual}. 
Additionally, enabling multiuser experiences that allow the users to collaborate and have their actions affect the \acp{VE} of others is an important goal~\cite{bachmann2019multi}.

% Future VR setups will eventually support multiple coexisting users fully immersed in mobility-wise unconstrained \acp{VE}.
% Such setups will be enabled by high frequency wireless networks primary operating in the millimeter Wave (mmWave) (i.e., 30-300 GHz) band~\cite{elbamby2018toward}. 
% For delivering real-time high-quality content to the likely mobile \ac{VR} users, the supporting wireless communication will have to be highly directional in both transmission and reception~\cite{liu2021learning}. 

In the future, \ac{VR} systems will have the capability to accommodate multiple users who can fully engage in immersive \acp{VE} without being limited by mobility. 
This advanced functionality will be made possible by high-frequency wireless networks primarily operating in the \acf{mmWave} band, ranging from 30 to 300 GHz~\cite{elbamby2018toward}. 
To provide mobile \ac{VR} users with high-quality content in real-time, the wireless communication that supports these systems will need to be highly directional in both transmission and reception~\cite{liu2021learning}.

% In transmission, directional mmWave beams will ``track'' the users’ movements for continuously maintaining \acf{LoS} connectivity with them.
% At the same time, \acf{RDW} is envisioned to be utilized for avoiding physical collisions among the users and between the users and the boundaries of the VR setups~\cite{bachmann2019multi}. 
% This will enhance users' immersion in \acp{VE} by allowing them to roam freely, while simultaneously imperceivably redirecting them in the physical setups for avoiding collisions.
% For supporting continuous \ac{LoS} maintenance, directional mmWave beams will have to be transmitted in a way that provides coverage at both current and near-future locations of the users.
% This motivates the need for users' short-term lateral movement prediction in full-immersive multiuser \ac{VR} setups with \ac{RDW}.

Directional \ac{mmWave} beams will follow users' movements during transmission to maintain \acf{LoS} connectivity with them. 
Meanwhile, \acf{RDW} will be utilized to prevent physical collisions between the users and \ac{VR} setup boundaries or other users~\cite{bachmann2019multi}. 
\ac{RDW} enables the users to explore \acp{VE} freely while subtly redirecting their physical movements for collision  avoidance, thus enhancing immersion. 
Short-term lateral movement prediction of the users can be used to support continuous \ac{LoS} connectivity, as directional \ac{mmWave} beams must provide coverage for both current and near-future user locations. 
This requirement highlights the importance of short-term lateral movement prediction in full-immersive multiuser \ac{VR} setups with \ac{RDW}.

% Short-term movement prediction for a natural human walk is an established topic in the community, which yielded \acp{RNN} in general and \acf{LSTM} networks as their particular instance suitable for this task (e.g.,~\cite{song2020pedestrian}).
% However, neither imperceivable nor perceivable resteering resemble a natural walk, suggesting the need for assessing the suitability of \acp{RNN} for predicting users' mobility under the constraints of \ac{RDW}.
% This topic received substantially less attention in the community.
% Nonetheless, the authors in~\cite{nescher2012analysis} have demonstrated that \acp{RNN}, specifically \ac{LSTM} networks, can be utilized for this purpose and feature encouraging prediction accuracy.

Predicting short-term movements in natural human walking is an established research topic, with \acf{LSTM} networks from the family of \acp{RNN} being particularly effective (e.g.,~\cite{song2020pedestrian}). 
Although these methods have proven useful for predicting natural walks, neither imperceptible nor perceptible resteering accurately mimics natural walking movements. 
This indicates a need to assess the suitability of \acp{RNN} in predicting \ac{VR} users' lateral mobility under the constraints of \ac{RDW}. 
Despite this, the topic has received relatively little attention in the community. 
Nevertheless, recent work by~\cite{nescher2012analysis} has demonstrated that \acp{RNN}, particularly \ac{LSTM} networks, can be applied for this purpose and feature promising levels of accuracy in a single-user setup.

% In this work, we extend these initial findings by considering information from \acp{VE} (i.e., the VR users' movement trajectory in a \ac{VE}) as an additional input feature for prediction, which is in contrast to the existing works that base the prediction solely on physical movement trajectories. 
% In our experimental evaluation, we consider a varying number of coexisting \ac{VR} users and different types of \acp{VE}.
% We show that the virtual movement trajectory provides additional useful context as it improves the prediction accuracy.

Our work builds upon previous research by incorporating context information from \acp{VE} into the prediction. 
Specifically, we utilize the users' movement trajectory in a \ac{VE} as an input feature, which differs from existing methods that solely rely on physical movement trajectories. 
Our experiments evaluate the impact of different numbers of coexisting users and types of \acp{VE} on the predictive accuracy. 
Our results demonstrate that incorporating virtual movement trajectory as an input context significantly enhances the accuracy of the prediction model.

% The users' real-world motions are envisioned to be accurately reflected within \acp{VE}, enabling them to seamlessly change their gaze direction.
% Flexible coverage is highly advantageous for receiver-side beamforming on a \ac{HMD} for supporting this requirement. 
% This is because a beam misalignment of even a few degrees can have a significant impact on the \ac{SNR}~\cite{abari2017enabling}.
% Hence, a flexible beam stretching in the head rotation direction can provide the \ac{HMD} with consistently high gain required for uninterrupted content delivery. 
% This approach can be used for supporting the reflection of user's motions on-screen within the motion-to-photon latency of 20~ms for avoiding nausea~\cite{elbamby2018toward}. 
% We argue that head rotations should be accurately predicted to proactively form such beams.
% Existing approaches for such predictions already feature encouraging accuracy, e.g.,~\cite{aykut2019realtime}.

It is envisioned that the users' real-world movements should be accurately reflected in \acp{VE}, allowing for seamless changes in gaze direction. 
To support this requirement, flexible coverage is highly advantageous for receiver-side beamforming on a \ac{HMD}. 
Even a slight beam misalignment can significantly affect the \ac{SNR}~\cite{abari2017enabling}, which is why a flexible beam stretching in the head rotation direction can provide the \ac{HMD} with the consistently high gain necessary for uninterrupted content delivery. 
This approach ensures that user motion is reflected on-screen within the motion-to-photon latency of 20~ms for avoiding nausea~\cite{elbamby2018toward}. 
We argue that accurate prediction of head rotations is needed to proactively form such beams. 
Existing approaches for such predictions are already highly accurate, as seen in e.g.,~\cite{aykut2019realtime}.

% Such prediction algorithms usually include \ac{DL} components converting users' orientational data to useful outputs. 
% Training, testing, and evaluating such algorithms requires massive amounts of orientational data.
% While these algorithms are the most ubiquitous consumers of orientational data, needs for substantial orientational data sources arise in other situations as well (e.g., receiver-side beamforming, enabling \ac{RDW}).
% Clearly, gathering these datasets is a costly labor-intensive process that does not scale well.
% A more efficient approach would be to apply synthetic data generation for augmenting existing datasets with new samples with insignificant changes to the distributions of the overall datasets. 

Typically, such prediction algorithms rely on \ac{DL} components for transforming the users' orientation data into valuable outputs. 
However, training, testing, and evaluating these algorithms necessitates collecting vast amounts of orientation data. 
While these algorithms are the primary consumers of orientation data, other situations also require large orientation data sources, such as receiver-side beamforming and enabling \ac{RDW}. 
Collecting these datasets is expensive and laborious, making it challenging to scale. 
Instead, a more effective approach would be to use synthetic data generation to supplement existing datasets with new samples, adding only minor variations to the overall dataset distribution.

% To the best of our knowledge, the literature on synthetic orientational data generation consists of an exploratory work basing such generation on the \ac{FFT}~\cite{blandino2021head}. 
% This approach considers time series of orientations as signals, which are converted to power spectral densities, followed by modelling the mean power spectral density.
% Perturbed versions are then converted back to orientational series, resulting in synthetic time series that closely match the mean of the input series. 
% In contrast, we use a TimeGAN approach from the family of \acp{GAN} for such generation. 
% We follow by experimentally demonstrating the model's beyond state-of-the-art capabilities.

To the best of our knowledge, there is currently only a single study on generating synthetic orientational data, exploring the use of \ac{FFT} for the generation~\cite{blandino2021head}. 
In this approach, the input orientation time series is treated as a signal and converted to power spectral densities, and the mean power spectral density is modeled. 
This method then generates synthetic time series by converting perturbed versions of the power spectral densities back to orientational series. 
This results in synthetic series that closely resemble the mean of the input ones. 
Our approach, on the other hand, uses a TimeGAN model from the family of \acp{GAN} for generating synthetic orientational data. We demonstrate experimentally the model's beyond state-of-the-art capabilities.

% In more general terms, with this work we argue in favor of utilizing predictive context-awareness for optimizing the performance of mmWave networks for supporting full-immersive multiuser VR with \ac{RDW}.
% We do that by showing that two context information instances, i.e., the VR users' lateral mobility under \ac{RDW} constraints and their orientational mobility, can be accurately predicted for a short timeframe.   
% We also discuss how these predictive context instances can intuitively be used for optimizing the performance of supporting mmWave networks along two dimensions, transmitter-side beamforming and beamsteering toward the VR users' \acp{HMD}, and receiver-side beamforming for coverage enhancements. 

In more general terms, this work advocates the utilization of predictive context-awareness to enhance the performance of \ac{mmWave} networks that support full-immersive multiuser \ac{VR} with \ac{RDW}. 
Specifically, we demonstrate that predictive context information, such as the users' lateral mobility under the constraints of \ac{RDW} and their orientational mobility, can be precisely forecasted for a brief period. 
Furthermore, we explore how these predictive context information instances can be utilized to optimize the performance of supporting mmWave networks. 
This optimization can occur along two dimensions: transmitter-side beamforming and beamsteering toward the \ac{VR} users' \acp{HMD}, and receiver-side beamforming for coverage enhancements.

%!TEX root = ieee_commag_main.tex
\section{System Overview}

% We consider a fully immersive multiuser \ac{VR} setup, as shown in Figure~\ref{fig:scenario}. 
% In particular, we consider a constrained physical environment in which fully immersive \ac{VR} setup is deployed. 
% The environment is constrained to create a safe perimeter for engaging in \acp{VE}. 
% Therefore, the only factors to consider as potential collision hazards from a VR user's perspective are the environment's boundaries and other users.

\begin{figure*}[!t]
\centering
\includegraphics[width=0.94\linewidth]{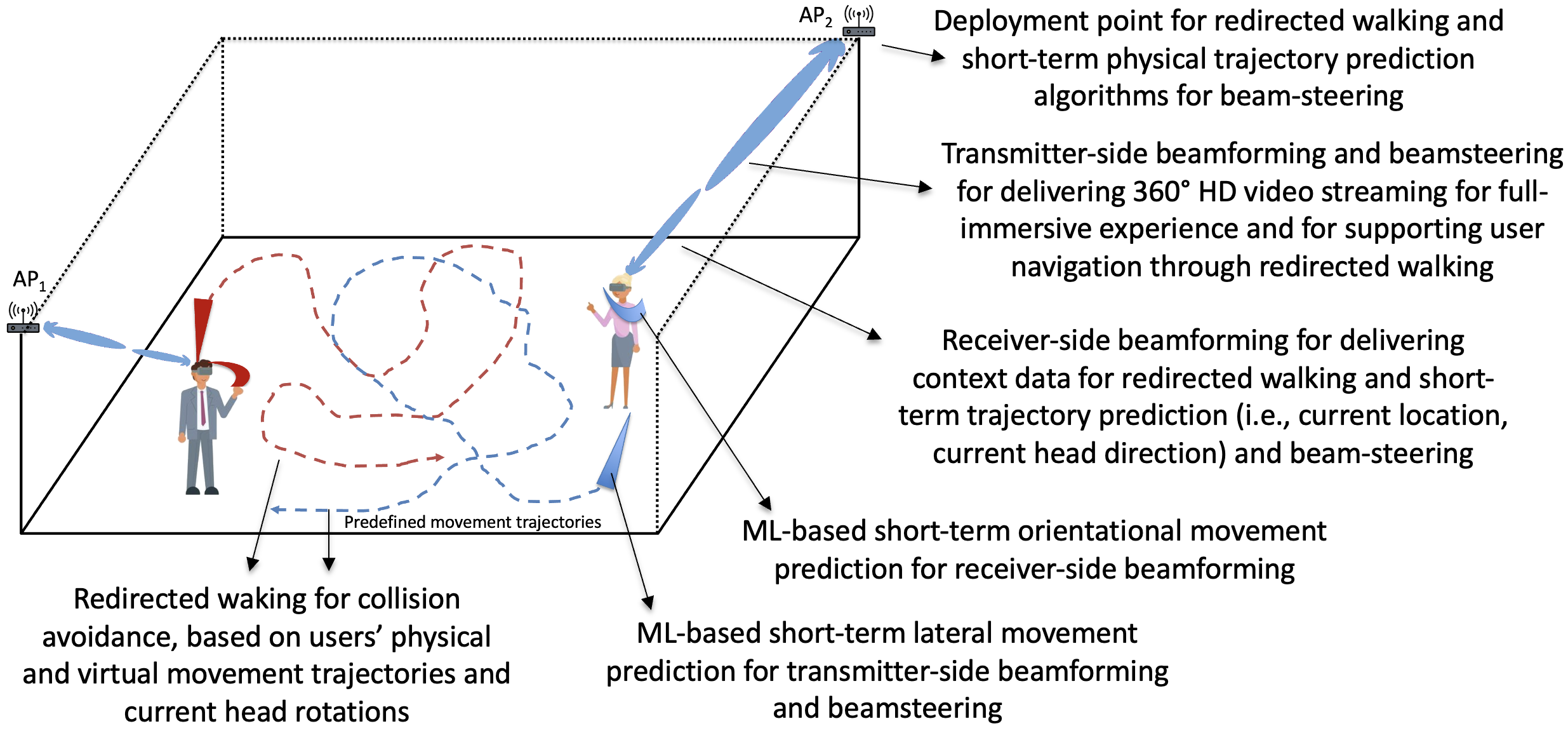}
\vspace{-1mm}
\caption{Considered Scenario for Full-Immersive
Multiuser Virtual Reality with Redirected Walking}
\label{fig:scenario}
\vspace{-3mm}
\end{figure*}

In Figure~\ref{fig:scenario}, we showcase a full-immersive multiuser \ac{VR} setup. 
Our focus lies on the deployment of this setup within a physically constrained environment that prioritizes user safety while engaging with \acp{VE}. 
This safe perimeter limits potential collision hazards for the users to the environmental boundaries and other users.

% \ac{RDW} is utilized for users' steering in a way that guarantees collision avoidance among them, as well as between them and the environmental boundaries.
% \ac{RDW} aims at enabling immersion of the users (i.e., allowing them to roam freely in potentially unconstrained \acp{VE}), while simultaneously and (ideally) imperceivably redirecting them in the physical space.

\ac{RDW} is employed to steer the users and ensure collision avoidance between them and the environmental boundaries, as well as among themselves. 
Its objective is to facilitate user immersion by enabling them to freely explore \acp{VE} without constraints, while seamlessly redirecting their movements in the physical space, (ideally) without causing any noticeable disruptions.

% This is accomplished through: i) curvature gains, i.e., rotations of a \ac{VE}, ii) translational gains, i.e., modifying linear movements in a \ac{VE} resulting in a change in the distance of the user’s physical travel, and iii) rotational gains, i.e., introducing additional rotations to the already rotating user. 
% A promising \ac{RDW} algorithm is \ac{APF}~\cite{bachmann2019multi}, which imperceivably re-steers the users into an open space by generating a force vector that is directed away from obstacles and has a length inversely proportional to its distance from each obstacle.
% The imperceivable resterring is enabled by abiding to the RDW noticeability thresholds empirically determined by Steinicke~\emph{et al.}~\cite{steinicke2009estimation}.
% When a collision is imminent and cannot be avoided, the resetting algorithm is triggered. 
% Artificial Potential Fields~-~Resetting (APF-R) is a resetting algorithm proposed in~\cite{bachmann2019multi} that uses the total force vector, calculated similarly to APF, to determine the angle the user has to physically turn toward. 
% This is followed by instructing a 2:1 turn. 
% In a 2:1 turn, the user's rotational speed gets increased, resulting in the user turning 360$^{\circ}$ in the VE while in reality only turning a smaller computed angle. 
% In this study, we utilize APF and APF-R for imperceivable resterring and resetting in the situations of an imminent collision, respectively.

The three ways of achieving imperceptible resterring in \acp{VE} are: i) curvature gains, which involve \ac{VE} rotations, ii) translational gains, which modify the users' linear movements to change their travel distances in \acp{VE}, and iii) rotational gains that introduce additional rotations to the already rotating users. 
A promising algorithm for achieving this is the \ac{APF}~\cite{bachmann2019multi}, which generates a force vector that guides the users away from obstacles and scales inversely with the distance from each obstacle, including other users. 
APF respects empirically determined \ac{RDW} noticeability thresholds~\cite{steinicke2009estimation}, resulting in imperceptible resterring. 
In case of an imminent collision, the resetting algorithm called \ac{APF-R}~\cite{bachmann2019multi}  is triggered. 
The \ac{APF-R} algorithm calculates the total force vector for determining the angle the users should physically turn toward, followed by instructing a 2:1 turn. 
During a 2:1 turn, the users' rotational speed increases, allowing them to turn 360$^{\circ}$ in the \ac{VE} while turning a smaller computed angle in reality. 
Our study uses \ac{APF} and \ac{APF-R} for imperceptible resterring and resetting in situations where a collision is imminent, respectively.

% VR contents are streamed to the users using highly directional mmWave communication.
% Specifically, highly directional beams are transmitted by an \ac{AP} in a way that they follow the movements of the users.
% This is done to continuously maintain \ac{LoS} connectivity with each user, resulting in a maximized link quality and enhanced \ac{QoE}.
% To do so, we envision the \acp{HMD} reporting their physical locations to the AP, which are then utilized for supporting both \ac{RDW} and beamsteering.
% Note that contemporary VR headsets such as the Oculus Quest 2 and Vive Cosmos already support the generation and provisioning of physical locations of the \acp{HMD} through on-board sensors and inside-out tracking.  

To ensure optimal \ac{QoE}, \ac{VR} content is delivered to the users via highly directional \ac{mmWave} communication. 
Specifically, an \ac{AP} transmits focused beams that track the users' movements, thereby continuously maintaining \ac{LoS} connectivity with each of them. 
This approach maximizes the link quality and enhances the users' \ac{QoE}. 
The process involves the \ac{VR} headset reporting its location to the \ac{AP}, which is then used to facilitate both \ac{RDW} and beamsteering. 
It is worth noting that modern \ac{VR} headsets, such as the Oculus Quest~2 and Vive Cosmos, already possess the capability to generate and share the physical locations of the \acp{HMD} via their built-in sensors and inside-out tracking.

% In terms of transmitter-side beamforming and beamsteering, the AP is envisioned to form beams that cover the users' current and near future locations, enabling \ac{LoS} maintenance during a current time instance, as well as in the near future.
% A number of such approaches have been discussed in the literature, one promising example being~\cite{lim2019beam}.
% On the receiver side, beamforming is envisioned to adapt in real-time to the users' head rotations, relying on the HMDs' built-in sensors providing accurate orientation estimates. 
% Interested readers can find more details on this aspect of the system in~\cite{struye2021millimeter}, where we present coVRage, a receiver beamforming solution in which, based on past and current orientations, the HMD predicts how the \ac{AoA} from the \ac{AP} will change and covers the \ac{AoA} trajectory with a dynamically shaped beam.  

The \ac{AP} aims at creating beams that cover the current and near-future locations of the users, facilitating \ac{LoS} maintenance at present and in the upcoming period. 
Several techniques, including~\cite{lim2019beam}, have been proposed in the literature to achieve this goal for transmitter-side beamforming and beamsteering. 
The beamforming on the receiver side is also expected to adapt to the users' head rotations using the \acp{HMD}' built-in sensors that provide accurate orientation estimates. 
For those interested in this system aspect, in~\cite{struye2021millimeter} we have introduced coVRage, a receiver beamforming technique that anticipates changes in the \ac{AoA} from the \ac{AP} using past and present orientations as references. 
Subsequently, the \ac{HMD} adjusts the beam dynamically to encompass the \ac{AoA} trajectory.
% For readers interested in this aspect of the system, in~\cite{struye2021millimeter} we introduce coVRage, a receiver beamforming solution that predicts the \ac{AoA} changes from the \ac{AP} based on past and current orientations. 
% The \ac{HMD} then dynamically shapes a beam to cover the trajectory of the \ac{AoA}. 

% Note that we consider as out of scope potential interruptions in the LoS connectivity between the AP and the users due to the obstructions caused by the other users.  
% An intuitive solution to the issue can be based on an AP handover if it is predicted that a user will block the LoS path of another one, again motivating the need for short-term prediction of the VR users' movements.    
% Alternative solutions based on \acp{IRS} have also been presented~\cite{liu2021learning}.

It should be noted that any possible interruptions in the \ac{LoS} connectivity between the \ac{AP} and its users caused by obstructions from other users are beyond the scope of our considerations. 
One solution to this issue could be to implement an \ac{AP} handover when there is a prediction that a user's movement will obstruct the \ac{LoS} path of another user. 
This emphasizes the need for accurately predicting \ac{VR} users' short-term movements. 
Additionally, solutions that utilize \acp{IRS} have also been proposed, e.g.,~\cite{liu2021learning}.
% An intuitive solution to this problem could involve an \ac{AP} handover if it is predicted that one user will block the \ac{LoS} path of another. 
% This reinforces the requirement for short-term prediction of \ac{VR} users' movements. 
% Alternative solutions based on \acp{IRS} have also been presented~\cite{liu2021learning}.

%!TEX root = ieee_commag_main.tex
\section{Toward Predictive Context-Awareness}

\subsection{Short-Term Lateral Movement Prediction}

% % Based on the discussion above, short-term prediction of the users' lateral movements will be utilized for both \ac{RDW} and transmitter-side beamsteering, while also being beneficial for \ac{LoS} obstruction avoidance.
% % It is intuitive that the accuracy of such prediction will affect the performance of other parts of the envisioned system, suggesting that its optimization will be beneficial from the perspective of the users' \ac{QoE}.  

The preceding discussion suggests that short-term prediction of users' horizontal movements will serve as a tool for \ac{RDW}, transmitter-side beamsteering, and \ac{LoS} obstruction avoidance. 
It is intuitive that the accuracy of such prediction will directly impact the effectiveness of other system components, making optimization crucial for improving the users' \ac{QoE}.

% % As mentioned before, \acp{RNN} are arguably the most suitable candidates for predicting the trajectory of lateral movements. 
% % \ac{RNN} is a class of artificial neural networks which contains multiple neurons of the same type, each passing a message to a succeeding one.
% % This allows RNNs to exhibit temporally dynamic behavior, making them suitable for tasks such as handwriting, speech recognition, and time-series prediction.
% % Due to its promise in predicting natural walking, we consider \ac{LSTM} as one of the most promising RNNs for predicting the trajectory of lateral movements under the constraints of \ac{RDW}.
% % A single neuron of an \ac{LSTM} network is depicted in Figure~\ref{fig:virtual_inputs}, together with the indications of its main building blocks.
% % In LSTM, a \emph{sigmoid} layer called the forget gate is utilized for deciding on the retainment of the previous cell state. 
% % A \emph{sigmoid} layer called the input gate jointly with a \emph{tanh} layer are utilized for updating the cell state.
% % The output is then based on the cell state filtered through a \emph{sigmoid} layer for deciding the part of the cell state to be outputted, normalized through a \emph{tanh} layer.

\begin{figure*}[!t]
% \vspace{-2m}
\centering
\includegraphics[width=\linewidth]{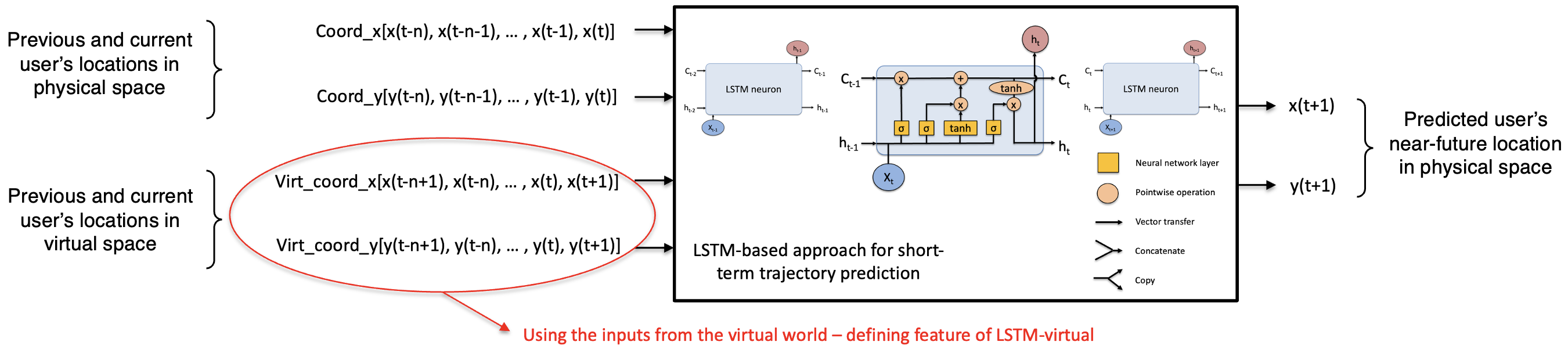}
% \vspace{-3mm}
\caption{Input features of the considered RNN approaches}
\label{fig:virtual_inputs}
\vspace{-3mm}
\end{figure*}

\acp{RNN} have been identified as the most appropriate choice for predicting lateral movements, as mentioned earlier. 
\acp{RNN} are a class of artificial neural networks that consist of multiple neurons of the same kind, each of which passes a message to a succeeding one. 
This enables \acp{RNN} to display dynamic behavior over time, making them well-suited for tasks such as speech recognition, handwriting, and time-series prediction. 
We consider \acp{LSTM} as one of the most promising \acp{RNN} for predicting the trajectory of lateral movements within the \ac{RDW} constraints, owing to their potential in predicting natural walking. 
Figure~\ref{fig:virtual_inputs} illustrates a neuron of an \ac{LSTM} network, as well as its key components. 
In \ac{LSTM}, a forget gate, which is a \emph{sigmoid} layer, is used to determine whether to retain the previous cell state. A \emph{sigmoid} layer called the input gate, together with a \emph{tanh} layer, is used to update the cell state. 
The output is then generated by filtering the cell state through a \emph{sigmoid} layer to determine which part of the cell state should be outputted, and normalizing it using a \emph{tanh} layer.

% % Existing \ac{LSTM}-based lateral movement predictors are predominantly based on previous physical locations and historical movements patterns of the users, taking their inspiration from predictions of natural walking trajectories.
% % However, in full-immersive VR setups with \ac{RDW} the users' movements are not fully natural, as they are continuously being re-steered for collision avoidance.
% % To do so, \ac{RDW} approaches direct the delivery of VR content in a way that points the users toward physical locations with no collision hazards. 
% % In other words, the redirections that are envisioned to occur in the \ac{VE} in the near future to avoid collisions are the defining feature of the mobility of VR users.

Most \ac{LSTM}-based lateral movement predictors rely on past physical locations and historical movement patterns to predict users' movements, taking inspiration from predictions of natural walking trajectories. 
However, in full-immersive \ac{VR} setups with \ac{RDW}, users' movements are not entirely natural. 
RDW techniques steer the users to avoid collisions by directing the delivery of \ac{VR} content towards collision-free physical locations. 
Therefore, the redirections that are expected to happen in the \ac{VE} to prevent collisions are a crucial aspect of \ac{VR} users' mobility.

% % Our intuition is that the \ac{RDW}-related inputs from the \acp{VE} are a valuable source of information for optimizing the near-future movement trajectory prediction for full-immersive VR users. 
% % To model such information, we define a notion of virtual locations of the users, i.e., their locations in the \acp{VE}.
% % We utilize a stream of historical virtual locations of the users as an input feature to the proposed \ac{LSTM} network, in addition to a stream of their historical physical locations. 
% % Based on the type of utilized input information, we distinguish the ``baseline'' and ``virtual'' versions of the approach. 
% % Note that, given that the decisions stemming from \ac{RDW}, and consequently the virtual coordinates of the users, are assumed to be known one step ahead in time compared to the physical coordinates, as shown in Figure~\ref{fig:virtual_inputs}.
% % We consider this to be a natural assumption, given that the virtual coordinates for the next time instance are in \ac{RDW} derived using the physical coordinates from a current time instance.

We argue that the \ac{RDW}-related inputs from the \acp{VE} can be a valuable source of information for optimizing near-future movement trajectory predictions for full-immersive \ac{VR} users. 
To incorporate this information, we introduce the concept of virtual locations, which represent the users' locations in the \acp{VE}. 
In addition to historical physical locations, we use a stream of historical virtual locations as input features to the proposed \ac{LSTM} network. 
Depending on the type of input information used, we distinguish between the ``baseline'' and ``virtual'' versions of our approach. 
It is worth noting that the virtual coordinates of the users are assumed to be known one time step ahead of their physical coordinates, as depicted in Figure~\ref{fig:virtual_inputs}. 
We consider this assumption to be natural, since the \ac{RDW}-derived virtual coordinates for the next time instance are based on the physical coordinates at current time.

\begin{figure*}[!t]
% \vspace{-2mm}
\centering
\includegraphics[width=0.75\linewidth]{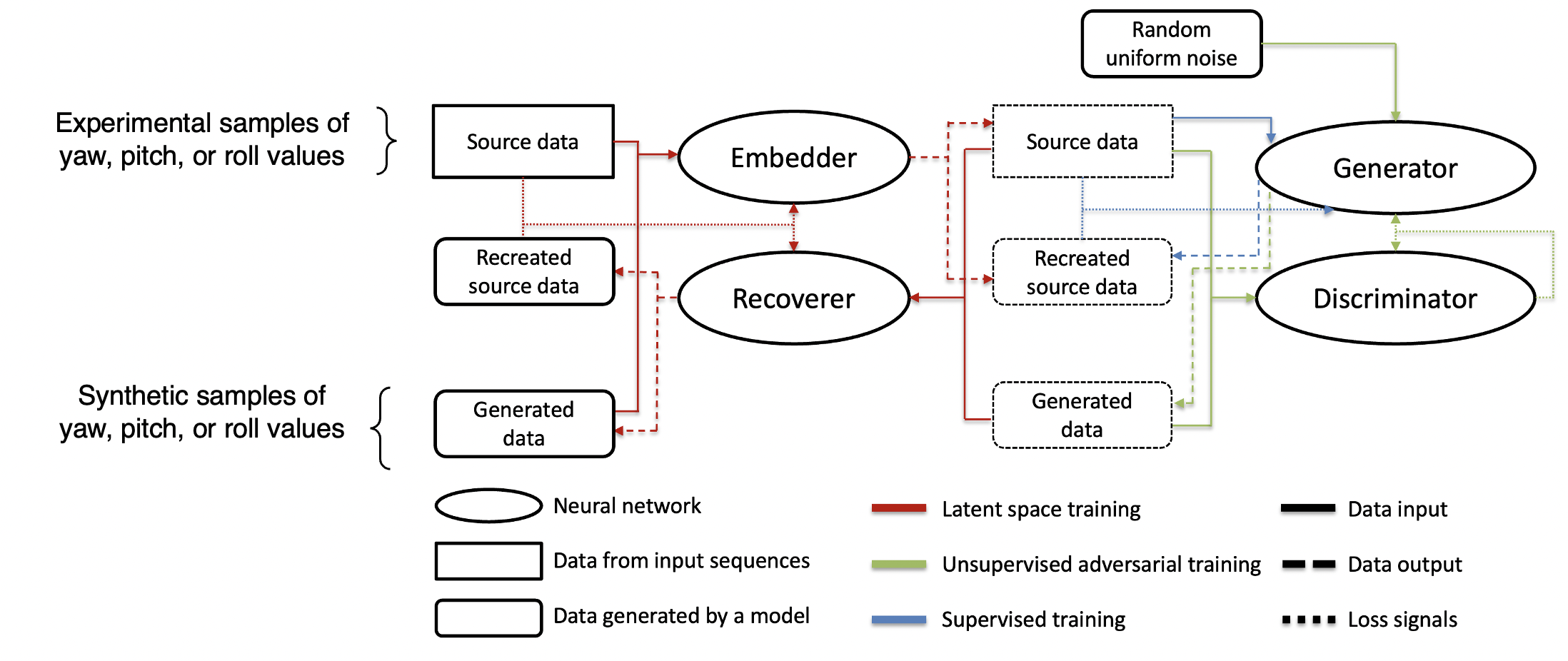}
\vspace{-1mm}
\caption{Illustration of the TimeGAN training process}
\label{fig:timegan}
\vspace{-2mm}
\end{figure*}

\subsection{Short-term Orientational Movement Prediction}

% The prediction of orientational movements assumes the availability of a reasonable amount of experimentally obtained inputs \textcolor{red}{(i.e., distributions of yaw, pitch, and roll movements)}. 
% For enabling existing predictors, there is a need for generating synthetic data based on the constrained experimental samples collected by the testers. 
% Current methods for such generation are primarily model-based, implying that the transfer of the generator across sources requires expert input and substantial modifications to its original design.
% Intuitively, a model-free approach featuring higher generalizability would be desired.
% \acp{GAN} represent a general \ac{DL} agent design, where its two sub-systems adversarially interact for generating samples that can be introduced to the original dataset without significantly affecting its overall distribution. 
% As such, \acp{GAN} are highly beneficial for developing model-free synthetic data generation approaches.

To predict orientational movements, a sufficient amount of experimentally obtained input data is required \textcolor{red}{(i.e., distributions of yaw, pitch, and roll movements)}. 
To make existing predictors more effective, synthetic data needs to be generated using the limited experimental data available. 
Current methods of generating synthetic data are mostly model based, which necessitates expert input and significant modifications to the original design to transfer the generator across sources. 
Ideally, a model-free approach with higher generalizability would be preferable. 
\acp{GAN} are a type of a general \ac{DL} agent design that involves two sub-systems interacting adversarially to generate samples that can be integrated into the original dataset without significantly altering its overall distribution. 
As a result, \acp{GAN} are extremely useful for developing model-free synthetic data generation techniques.

% In a \ac{GAN}-based system, the generator initially utilizes random noise for generating synthetic data.
% The discriminator is a supervised learning-based classifier that categorizes presented samples as real (i.e., originating from the source dataset) or fake (i.e., originating from the generator). 
% The generator is unable to access the source samples, but only the discriminator's loss function.
% Thus, it can aim at optimizing the loss and eventually adjusting its outputs.
% This interactive procedure pushes the generator to produce more realistic samples closely matching the original distribution, in turn discovering more subtle contrasts between real and synthetic samples.

The \ac{GAN}-based system starts with the generator utilizing random noise to create synthetic samples. 
Meanwhile, the discriminator acts as a supervised learning-based classifier, identifying whether a presented sample is genuine (coming from the source dataset) or fake (coming from the generator). 
The generator cannot access the source samples and it only has access to the discriminator's loss function. 
Therefore, the generator focuses on optimizing the loss to improve its output. 
This iterative process drives the generator to generate increasingly realistic synthetic samples that resemble the original distribution, unveiling more nuanced differences between real and synthetic samples.

% There is a need to maintain correlation between the samples as they represent a time series. 
% We propose the utilization of TimeGANs for generating synthetic orientational datasets due to their intrinsic suitability in maintaining time-correlation within a series. 
% The proposed design of the TimeGAN training \textcolor{red}{process} is depicted in Figure~\ref{fig:timegan}.
% The discriminator and generator consist of \acp{GRU} capable of dealing with time-dependencies.
% We introduce the embedder and recoverer subsystems for data encoding to a latent space of lower dimensionality, as well as decoding from the latent to the original dimension. 
% These two subsystems are initially trained, followed by the generator producing samples in the latent space.
% The samples are then converted to time series during the recovery process.
% Finally, through supervised learning the generator generates complete latent representations of incomplete series using the source dataset.
% The loss function is further encouraging the generator to capture the time-correlation within the series by representing the distance between the synthetically generated data in a given time step and actual source data at similar time steps. 
% We alternatingly use adversarial and supervised learning with their loss functions implemented as cross-entropy and \ac{MSE} losses, respectively. 

There is a need to maintain correlation between the samples as they represent a time series.
We suggest using TimeGANs to create synthetic orientational datasets, as they are particularly well-suited to this task due to their ability to maintain time-dependencies. 
To achieve this, our proposed TimeGAN training \textcolor{red}{process} consists of a discriminator and generator, both of which use \acp{GRU} to handle time-dependencies. 
Additionally, we introduce an embedder and recoverer subsystem for data encoding into a lower-dimensional latent space, followed by decoding back to the original dimension. 
We first train these two subsystems before the generator produces samples in the latent space, which are then converted to time series through the recovery process. 
To generate complete latent representations of incomplete series from the source dataset, we use supervised learning. 
The generator is encouraged to capture time-correlation within the series through a loss function that measures the distance between synthetically generated data and the actual source data at similar time steps. 
To achieve this, we alternate between adversarial and supervised learning, using cross-entropy and \ac{MSE} losses, respectively. 
Please refer to Figure~\ref{fig:timegan} for a visual representation of our proposed design.

%!TEX root = ieee_commag_main.tex

\begin{figure*}[!t]
% \vspace{-2mm}
\centering
\includegraphics[width=0.80\linewidth]{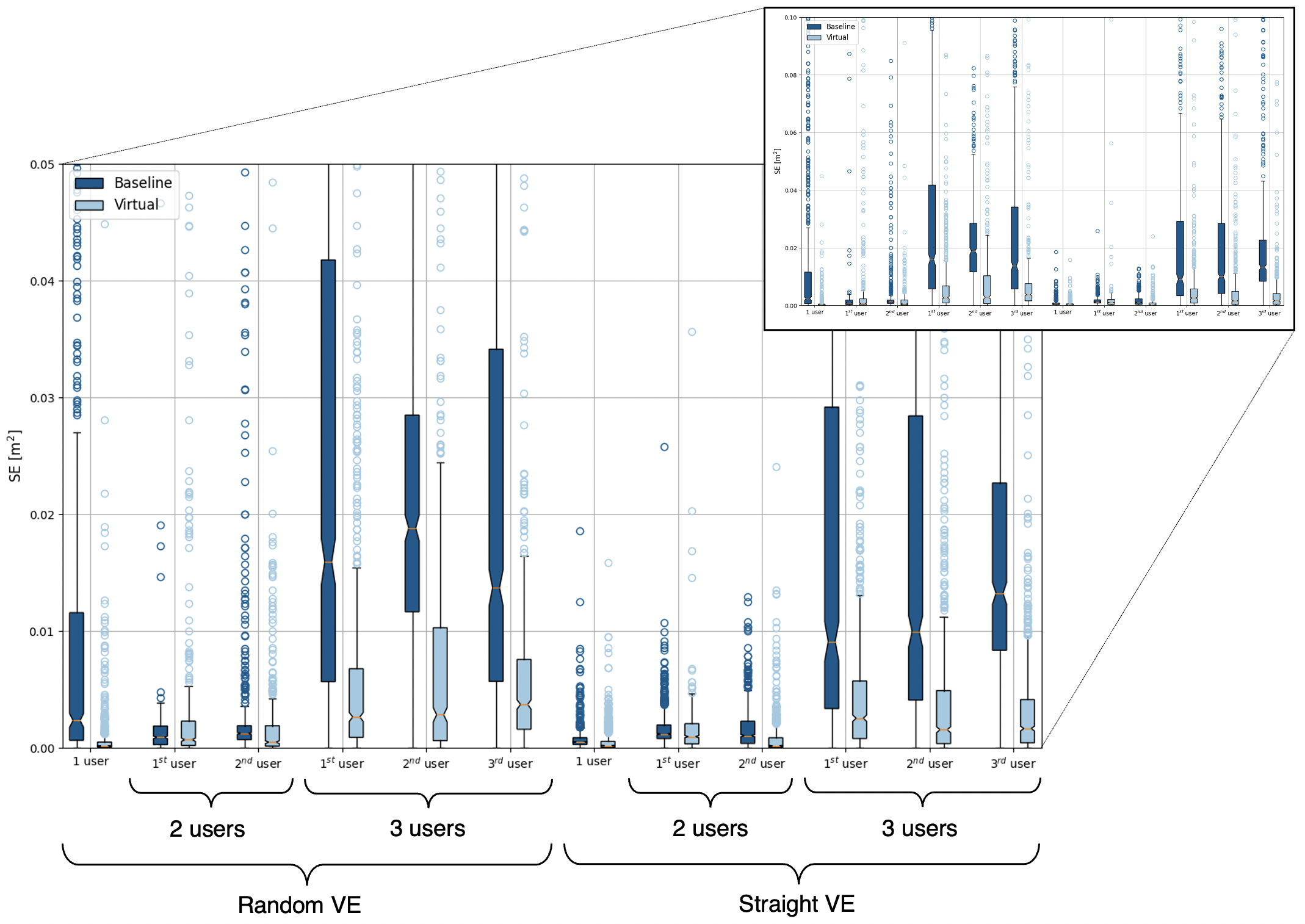}
% \vspace{-1mm}
\caption{Squared errors (SEs) achieved by different versions of the LSTM approach}
\label{fig:mse_lstm}
% \vspace{-2mm}
\end{figure*} 

\section{Evaluation Setup and Results}

We assess the performance of LSTM-based short-term lateral movements prediction in potentially multiuser VR setups with \ac{RDW}.
This is followed by the performance assessment of TimeGAN for generating synthetic orientational movement datasets, envisaged as a primer for predicting orientational movements of fully immersed VR users.

\subsection{Short-term Lateral Mobility Prediction}

% The deployment environment is defined as a square with sizes of 15~m. 
% This sizes of the environment were determined experimentally with the aim of minimizing the environmental size required for maintaining acceptable level of noticeability of a \ac{VE}, while simultaneously abiding to the practical limitations of the future deployment sites (e.g., in residences).
% To accommodate such an environment (by precisely measuring and marking the environmental boundaries) and guarantee no collision with environmental obstacles for the testers, we utilized a 106$\times$65~m$^2$ outdoor field near the University of Antwerp. 
% In the setup, the server that was running the \ac{RDW} algorithm (i.e., APF) was a Windows~10-based MSI GS66 laptop with an Intel i7 processor, 16~GB of RAM and WiFi~6. 
% The utilized HMDs were Android-based Oculus Quests~2 with Qualcomm Snapdragon XR, 120~Hz refresh rate and 6~GB of RAM. 
% A wireless hotspot using a Samsung Galaxy S8 was providing connectivity between the server and the HMDs.

The deployment environment utilized in the experiment was a square with dimensions of 15 meters. 
These dimensions were determined through experimentation to find the optimal size that would ensure maintaining acceptable level of noticeability, while still being practical for future deployment (e.g., in residential settings). 
To create this environment and prevent testers from colliding with obstacles, a 106$\times$65~m$^2$ outdoor field near the University of Antwerp was utilized. 
The server running the \ac{RDW} algorithm (i.e., APF) was a Windows 10-based MSI GS66 laptop with an Intel i7 processor, 16~GB of RAM, and WiFi~6. 
The \acp{HMD} used were Android-based Oculus Quests2 with Qualcomm Snapdragon XR, a 120~Hz refresh rate, and 6~GB of RAM. Connectivity between the server and the \acp{HMD} was provided by a wireless hotspot using a Samsung Galaxy S8.

% We designed two VEs in Unity to assess the accuracy of prediction. 
% In the ``straight path'' experience, the testers were instructed to follow a straight path during the full duration of the VE. 
% This was considered as the worst-case scenario in terms of the noticeability and performance of the RDW algorithm. 
% In the ``random path'' experience, the testers walked in an open environment and were encouraged to follow a randomly curved path. 
% Hence, the curvature introduced by the \ac{VE} itself was expected to benefit the RDW algorithm and be less noticeable than the straight path alternative. 

Two \acp{VE} were designed in Unity to evaluate the prediction accuracy. 
In the ``straight path'' experience, the testers were instructed to follow a straight path throughout the \ac{VE}, representing the worst-case scenario for the noticeability and performance of the \ac{RDW} algorithm. 
In the ``random path'' experience, the testers were encouraged to follow a randomly curved path in an open environment. 
The expectation was that the curved path introduced by the \ac{VE} would be less noticeable and benefit the \ac{RDW} algorithm compared to the straight path.

% Conceptually, the experiments consisted of the testers walking in an unbound VE while being confined to a restricted physical environment. 
% The positional data was sent from the HMDs to the server, where the RDW algorithm steered the testers within the confined physical environment to avoid collisions with other testers and environmental borders. 
% In each experiment, we were assisted by up to \textcolor{red}{three} testers coexisting in the environment and fully immersed in a \ac{VE}.
% At the start of each experiment, the testers were instructed to simply follow a predefined path and enjoy the experience. 
% They were also informed that a reset could happen due to the recommendations from the RDW algorithm (i.e., APF-R). 
% In case of the reset, the tester would see a stop sign followed by the world rotating and guiding them in a recommended direction. 
% The duration of each experiment was set to 5 minutes because, as there are no distractions and interactions in the VEs, the testers would loose interest soon afterwards.

The experiments involved the testers walking in an unbounded \ac{VE} while being physically confined to a restricted environment. The positional data from the \acp{HMD} was sent to the server, where the \ac{RDW} algorithm guided the testers within the physical boundaries to avoid collisions with other testers and environmental borders. 
Each experiment involved up to three testers coexisting in the environment and fully immersed in the \ac{VE}. 
At the beginning of each experiment, the testers were instructed to follow a predefined path. 
They were informed that a reset might occur based on recommendations from the RDW algorithm (i.e., APF-R). 
If a reset occurs during the testing process, the testers would encounter a stop sign, followed by the \ac{VE} rotating to provide guidance in the suggested direction. 
To maintain engagement, the duration of each experiment was limited to 5 minutes because the testers would lose interest afterwards as there were no distractions nor interactions within the VEs.

Figure~\ref{fig:mse_lstm} illustrates the performance of the LSTM-based prediction using \acp{SE} as the metric of interest. 
Comparing the baseline with the version that incorporates virtual coordinates alongside physical ones, it is evident that the latter generally outperforms the former. 
In a single-user system and for both VEs, the average per-user SE of the prediction is reduced from around 0.001~m$^2$ in the baseline to less than 0.0005~m$^2$ when utilizing additional context from the \ac{VE}, resulting in a twofold increase in prediction accuracy. 

The improvements become more significant with an increasing number of users. 
For instance, in a three-user system, utilizing context instances from \acp{VE} leads to improvements of up to 75\% in the worst case, as depicted in the figure. 
These findings demonstrate the potential of leveraging specific context information from \acp{VE}, such as virtual coordinates of the users, to enhance the performance of \ac{LSTM}-based predictors of near-term lateral movements. 
Moreover, the introduction of a second user does not notably affect prediction accuracy in the considered \acp{VE}, indicating the usefulness of short-term lateral movement prediction in \textit{multiuser} \ac{VR} systems with \ac{RDW}. 
However, the accuracy decreases considerably when a third tester is introduced, regardless of the \ac{VE} type. 
This observation highlights the importance of appropriately sizing the physical environment based on the number of users and their mobility patterns. 
Notably, the prediction accuracy in the straight path \ac{VE}, which is the worst-case scenario for \ac{RDW}, is better than that in the random path \ac{VE}. 
This is likely because the straight path \ac{VE} introduces less curvature in the testers' movements, and these more linear movements can be more accurately predicted.

\begin{figure*}[!t]
\centering
\includegraphics[width=0.72\linewidth]{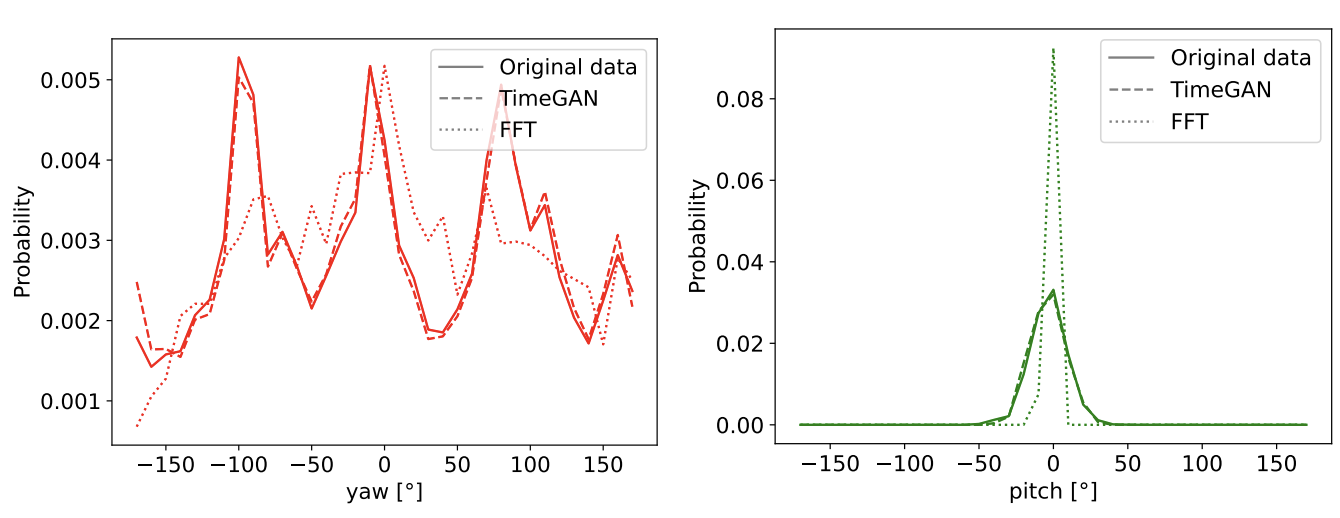}
\vspace{-3mm}
\caption{Distribution of Yaw and Pitch values across all time steps of all samples}
\label{fig:timegan_results}
\vspace{-4mm}
\end{figure*}

\subsection{Training of Orientational Mobility Predictors}

% In an immersive \ac{VE} from~\cite{chakareski20206dof} three testers navigated freely during six two minute-long sessions.  
% During that time, their full poses (i.e., lateral and orientational movements) were sampled at the frequency of 250~Hz. 
% We utilize the orientational traces provided in yaw-pitch-roll format from this dataset for assessing the performance of our approach for generation of synthetic orientational datasets. 
% Specifically, we quantized the \acp{PDF} of the yaw, pitch, and roll movements into 10$^{\circ}$-wide buckets centered around 0$^{\circ}$.
% The resulting \ac{PDF} of the yaw motion is substantially more complex in comparison to the other two movement types.
% This is a result of the points of interest being spread across the horizontal plane of the testers.
% In particular, the users' gaze was mostly directed toward one of the walls as the \ac{VE} was placed indoors, explaining the local maxima in Figure~\ref{fig:timegan_results} observed around -90, 0, and 90$^{\circ}$. 

During six two-minute-long sessions in an immersive \ac{VE} from~\cite{chakareski20206dof}, three testers were free to navigate and have their full poses, including lateral and orientational movements, sampled at a frequency of 250~Hz. 
The orientational traces from this dataset were used to evaluate our approach for generating synthetic head rotation datasets, with a focus on the \acp{PDF} of the yaw, pitch, and roll movements. 
We quantized the \acp{PDF} into 10$^{\circ}$-wide buckets centered around 0$^{\circ}$. 
The yaw motion \ac{PDF} was found to be substantially more complex than the other two types of movements, presumably due to the points of interest being spread across the testers' horizontal plane. 
As the \ac{VE} was placed indoors, the users' gaze was primarily directed toward one of the walls, which explains the local maxima observed in Figure~\ref{fig:timegan_results} around -90, 0, and 90$^{\circ}$.

% The pitch and roll samples feature normal distributions that are easily generated by neural networks.
% The distribution of yaw samples features multiple peaks and discontinuities over time due to the samples being constrained within -180 and 180$^{\circ}$, thus its generation represents a more complex challenge.
% We utilize a quantile transformer for non-linearly transforming the data to mitigate its non-normality. 
% To avoid discontinuity, we shift the remainder of a series by 360$^{\circ}$ upon observing a discontinuity. 
% These transformations imply that the data range cannot be established in advance anymore, but the samples remain within practically useful boundaries. 
% The introduced transformations are reversible, hence synthetic data can be back-transformed to original representations.

The normal distributions of the pitch and roll samples are easily reproduced by neural networks. 
However, the distribution of yaw samples is more complex due to the multiple peaks and discontinuities that arise over time, since these samples are limited to the range between -180 and 180$^{\circ}$. 
To overcome this challenge, we employ a quantile transformer to non-linearly transform the data and address its non-normality. 
In addition, we shift the remainder of a time series by 360$^{\circ}$ to avoid discontinuities. 
These transformations mean that the data range cannot be predetermined anymore, but the samples still fall within practically useful boundaries. 
Furthermore, these transformations are reversible, enabling synthetic data to be backtransformed to their original representations.

% GAN training time and complexity scale with the input data size.
% Given that \ac{DL} requires large quantity of distinct samples, using a sliding window we subdivided each series into 25 samples-long instances. 
% We also down-sampled the data, thereby not significantly reducing its utility. 
% Intuitively, a human can perform a small number of distinct head rotations in a short time-frame. 
% These motions will be relatively smooth as implied by the law of inertia, hence they can be accurately recreated using a simple interpolation. 
% Given that the energy is focused around the lowest frequencies, with over 90\% below 5~Hz, according to the Shannon-Nyquist sampling theorem this information will be maintained when downsampling to 10~Hz. 

The time and complexity of \ac{GAN} training increase in proportion to the input data size. 
To accommodate the demand for a large number of distinct samples in GAN-based \ac{DL}, we divided each time series into instances of 25 samples using a sliding window. 
We also applied data downsampling, which did not significantly impact its utility. 
Based on the law of inertia, humans can perform a limited number of distinct head rotations within a brief timeframe, resulting in relatively smooth motions that can be accurately reconstructed using simple interpolation. 
With most of the energy concentrated in the lowest frequencies, over 90\% below 5 Hz, the Shannon-Nyquist sampling theorem confirms that this information can be preserved by downsampling to 10~Hz.

The dataset is, therefore, divided into 23,700 samples of 25 points, each sample being 1.5~s long. 
The data was provided to TimeGAN for fitting the quantile transformer, followed by generating a synthetic dataset 10 times larger than the original one, once every 10 epochs. 
This was done because GANs are challenging to train and known to significantly degrade when overtrained.
The procedure yielded a well-distributed synthetic dataset, despite the fact that we did not optimize the epoch hyperparameter nor repeat the resource intensive training procedure.
The generated dataset is the final result of the system, therefore the system will not be presented with inputs of slightly varied distributions later on.
This is because the input comes from samples of uniformly distributed noise, implying that data overfitting can be excluded.
Finally, we generated a baseline synthetic dataset using the \ac{FFT} approach from~\cite{blandino2021head}. 
The result is a 30,000 steps-long series, which we further downsample and divide into shorter samples using the above-discussed sliding window approach.

% The distributions of yaw and pitch values are presented in Figure~\ref{fig:timegan_results}. 
% The roll distribution is omitted due to it being range-constrained as it captures uncomfortable head tilting.
% For this reason, both FFT and TimeGAN-generated synthetic datasets match it closely. 
% The pitch's distribution is slightly wider and here \ac{FFT} fails to match the distribution, in contrast to TimeGAN. 
% With the yaw distribution this behavior is further pronounced. 
% Specifically, its three local maxima are closely matched by TimeGAN and significantly less so by the \ac{FFT} approach.
% \textcolor{red}{For example, the Kullback-Leibler divergence of the target yaw distribution is 0.00235 when compared with the TimeGAN-generated one, and 0.0447 (i.e., almost 19 times higher) when compared to FFT-resulting one.}
% The results can be further strengthened by the fact that only FFT was hand-crafted to match this distribution.
% We argue the results demonstrate the promise of utilizing TimeGANs for generating synthetic head rotation datasets.

Figure~\ref{fig:timegan_results} presents the distributions of yaw and pitch values. 
The roll distribution has been omitted because it is range-constrained as it captures uncomfortable head tilting. 
Hence, both the \ac{FFT}- and TimeGAN-generated synthetic datasets closely match the roll distribution. 
However, the pitch distribution is slightly wider, and the \ac{FFT} approach already fails to match it accurately, in contrast to TimeGAN. 
This trend is further pronounced in the yaw distribution, where TimeGAN closely matches its three local maxima, while \ac{FFT} does so to a lesser extent. 
\textcolor{red}{For example, the Kullback-Leibler divergence of the target yaw distribution compared to the TimeGAN-generated one is 0.00235, while the same compared to the FFT-resulting distribution is 0.0447 (i.e., almost 19 times higher).} 
It is also worth noting that only FFT was hand-crafted to match this distribution. 
Based on these results, we argue that utilizing TimeGANs for generating synthetic head rotation datasets shows great promise.

% \vspace{-2mm}
%!TEX root = ieee_commag_main.tex
% \vspace{-1mm}
\section{Conclusion}
\label{sec:conclusion}

% We have shown that \acf{LSTM}-based \acfp{RNN} are a promising candidate for near-future lateral movement prediction in multiuser full-immersive \acf{VR} with \acf{RDW}.
% We have also demonstrated the benefits of utilizing virtual context such as the movement trajectory in a \acf{VE} as an input feature for the prediction.
Our research has demonstrated that \acf{LSTM}-based \acfp{RNN} hold promise for accurately predicting lateral movements in multiuser full-immersive \acf{VR} environments with \acf{RDW}. 
Additionally, we have showcased the advantages of incorporating virtual context, such as the movement trajectory within a \acf{VE}, as an input feature for enhancing the prediction accuracy.
Moreover, we have proposed a TimeGAN-based approach for generating synthetic head rotation data, envisioned to serve as a primer for training orientational movement predictors in full-immersive \ac{VR} setups.
On a high level, we advocate for the utilization of predictive context-awareness in optimizing the connectivity in next generation VR setups. 
We expect this approach to be beneficial in applications ranging from dynamic multimedia encoding to millimeter Wave (mmWave) beamforming.

Note that we did not optimize the duration of the prediction window, but this duration should intuitively depend on the transmitter-side beamforming and beamsteering operating in the 100~ms timeframe considered in this work. 
Deriving optimal hyperparameterizations of the presented approaches was also not in scope, as the goal was to demonstrate the feasibility of predictive context-awareness.
We consider addressing these limitations as a part of our future efforts. 
We argue that other context instances, for example the users' full \ac{3D} pose estimates, might be of interest in future VR systems, e.g., for enabling touch-like feedback or mobility-wise unconstrained portrayal of users in \acp{VE}.

\section*{Acknowledgments}
This work was supported by the MCIN / AEI / 10.13039 / 501100011033 / FEDER / UE HoloMit 2.0 (nr. PID2021-126551OB-C21) /  UNICO-5G I+D Open6G (TSI-063000-2021-6). The work was also funded by the Research Foundation - Flanders (nr. G034322N and 1SB0719N).

\renewcommand{\bibfont}{\footnotesize}
\printbibliography

\newpage

\section*{Biographies}

\vspace{-53pt}
\begin{IEEEbiography}[{\includegraphics[width=1in,height=1.25in,clip,keepaspectratio]{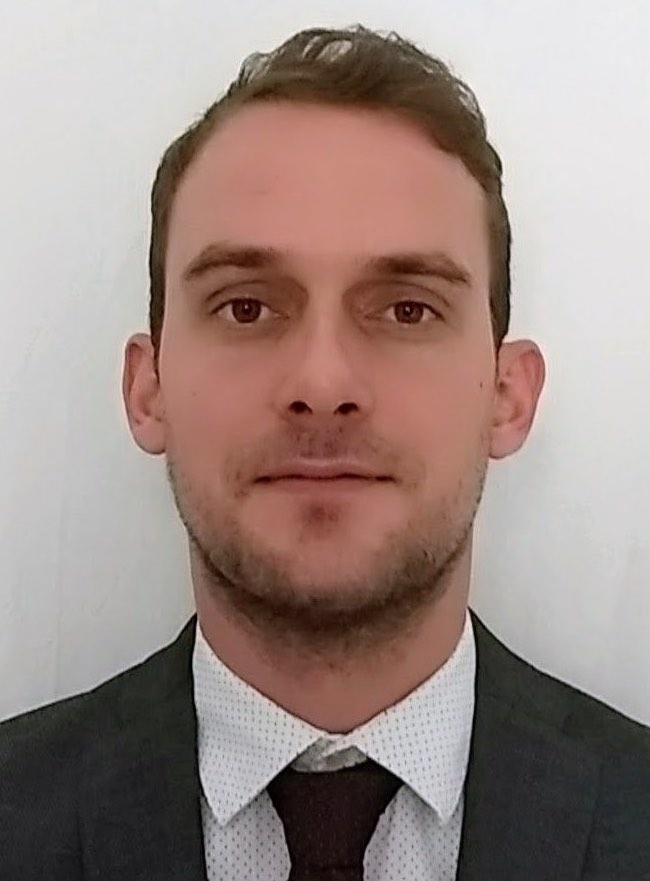}}]{Filip Lemic} is a senior researcher at the i2Cat Foundation. He held positions at the University of Antwerp, imec, Universitat Politècnica de Catalunya, University of California at Berkeley, Shanghai Jiao Tong University, FIWARE Foundation, and Technische Universität Berlin. He received his M.Sc. and Ph.D. from the University of Zagreb and Technische Universität Berlin, respectively. 
\vspace{-20pt}
\end{IEEEbiography}

\begin{IEEEbiography}[{\includegraphics[width=1in,height=1.25in,clip,keepaspectratio]{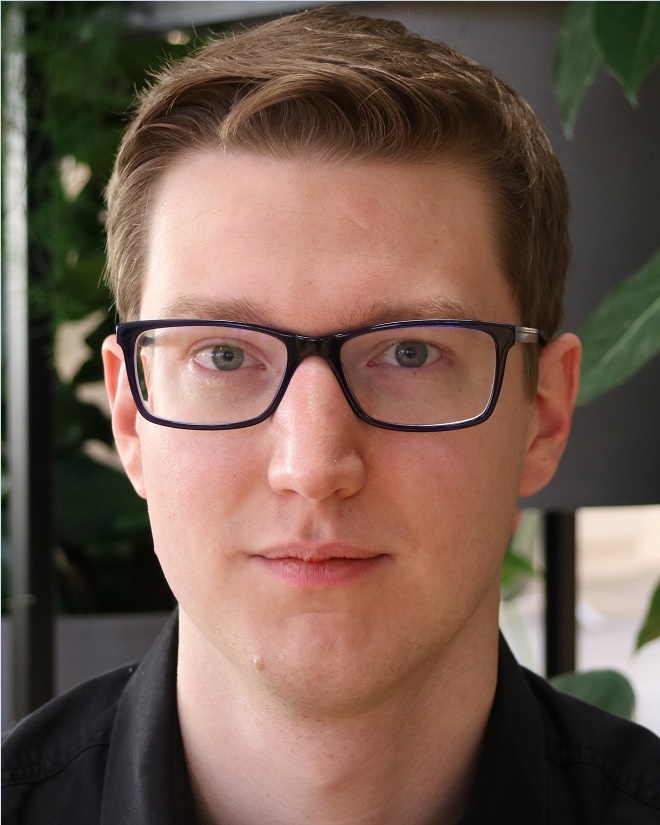}}]{Jakob Struye} is a Ph.D. researcher in the field of wireless networking at the IDLab research group (University of Antwerp) and imec research institute, Belgium. He obtained his B.Sc (2015) and M.Sc. (2017) in Computer Science at the University of Antwerp. His current research focuses on leveraging extremely high frequency wireless networks in the millimeter wave bands to improve the performance of truly wireless Virtual and Augmented Reality experiences, and he has experience in applying Artificial Intelligence to time series prediction problems. 
\vspace{-23pt}
\end{IEEEbiography}

\begin{IEEEbiography}[{\includegraphics[width=1in,height=1.25in,clip,keepaspectratio]{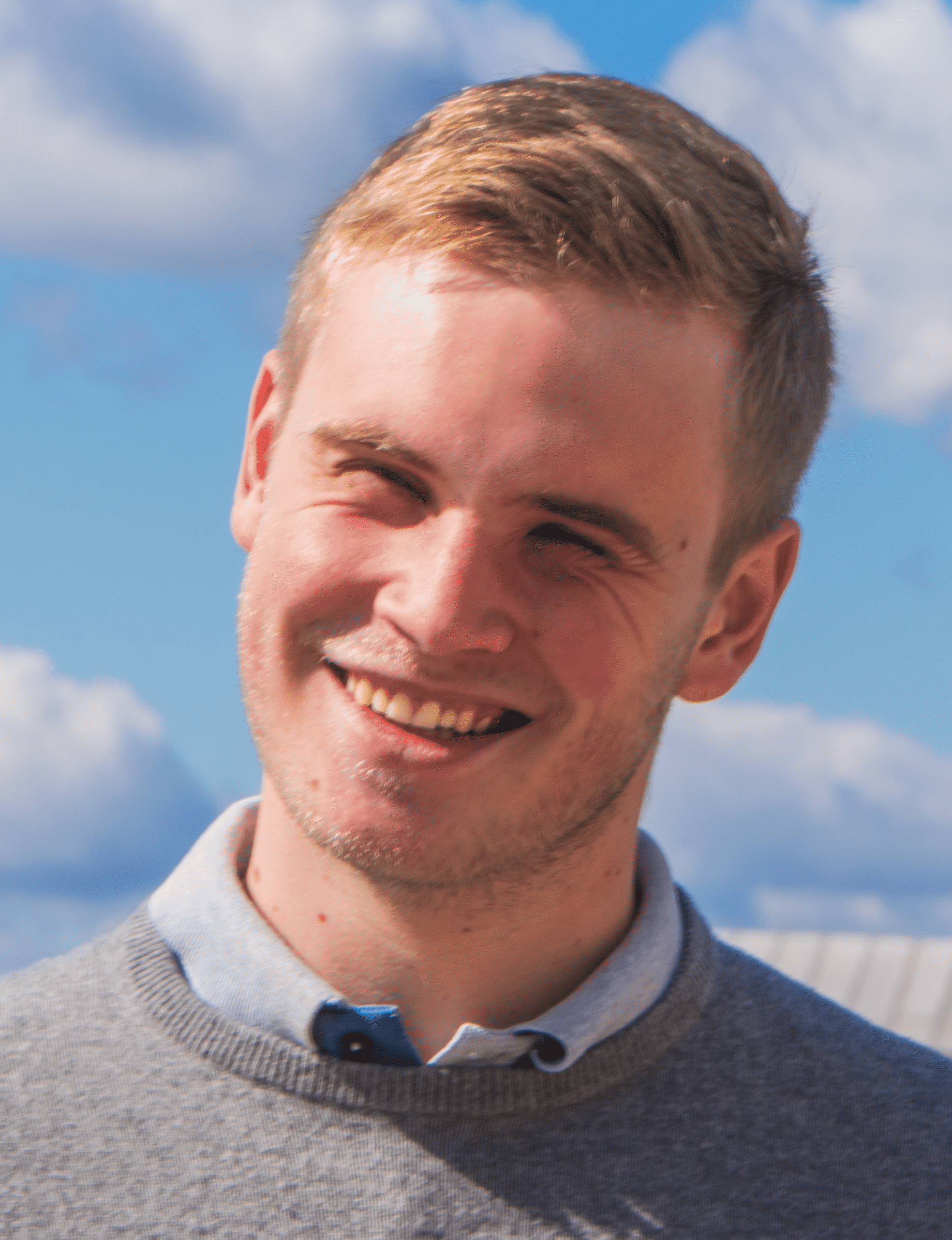}}]{Thomas Van Onsem} is a graduated M.Sc. in software engineering from the University of Antwerp. He has an interest in influential and impactful technologies. Currently, he assists companies' digital transformations as a technical consultant for Exellys.
\vspace{-20pt}
\end{IEEEbiography}

\begin{IEEEbiography}[{\includegraphics[width=1in,height=1.25in,clip,keepaspectratio]{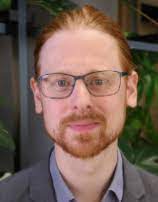}}]{Jeroen Famaey} is a research professor at IDLab research group of the University of Antwerp and imec, Belgium. He leads a team of researchers at IDLab, focusing on future wireless network technologies and protocols. His research has led to the publication of over 160 peer-reviewed journal articles and conference papers, and 7 granted patents.
\vspace{-20pt}
\end{IEEEbiography}

\begin{IEEEbiography}[{\includegraphics[width=1in,height=1.25in,clip,keepaspectratio]{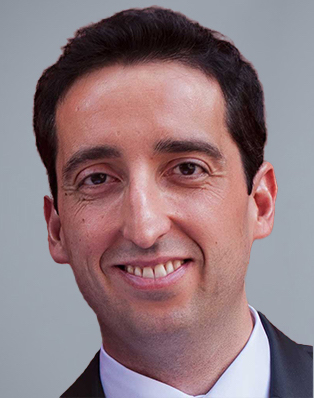}}]{Xavier Costa Perez}
is a research professor at ICREA, scientific director at i2Cat and head of 5G/6G R\&D at NEC Laboratories Europe. He received his M.Sc. and Ph.D. degrees from the Polytechnic University of Catalonia and was the recipient of a national award for his Ph.D. thesis.
\vspace{-10pt}
\end{IEEEbiography}



% \vfill

\end{document}